\documentclass[12pt,]{article}
\usepackage{graphics}
\usepackage{caption2}
\def\beq{\begin{equation}}
\def\eeq{\end{equation}}
\def\bea{\begin{eqnarray}}
\def\eea{\end{eqnarray}}
\def\bq{\begin{quote}}
\def\eq{\end{quote}}

\def\CMP{{\it Commun.Math.Phys.} }

\def\NC{{\it Nuovo Cimento} }
\def\NP{{\it Nucl.Phys.} }
\def\PL{{\it Phys.Lett.} }
\def\PR{{\it Phys.Rev.} }
\def\PRL{{\it Phys.Rev.Lett.} }
\def\PRTS{{\it Physics Reports} }

\def\gappeq{\mathrel{\rlap {\raise.5ex\hbox{$>$}}
{\lower.5ex\hbox{$\sim$}}}}

\def\lappeq{\mathrel{\rlap{\raise.5ex\hbox{$<$}}
{\lower.5ex\hbox{$\sim$}}}}

\def\Toprel#1\over#2{\mathrel{\mathop{#2}\limits^{#1}}}

\def \><{^{>}_{<}}
\def \<>{^{<}_{>}}

\begin{document}
\pagestyle{empty}
\setcounter{page}{1}

\begin{flushright}
{CERN-PH-TH/2007-069}\\
\end{flushright}
\vspace*{5mm}
\begin{center}
{\bf STATUS OF THE HEAVY QUARK SYSTEMS} \\
\vspace*{1cm}
{\bf Andr\'e Martin} \\
\vspace{0.3cm}
Theoretical Physics Division, CERN \\
CH - 1211 Geneva 23 \\ 
\vspace*{4cm}
{\bf ABSTRACT} \\ 
\end{center}
\vspace*{1cm}
\noindent
We review various inequalities on the order and the spacing of energy levels, wave function at the origin, etc... which were obtained since 1977 in the framework of the Schr\"odinger equation and applied to quarkonium and also to muonic atoms and alcaline atoms.  We also present a fit of mesons and baryons made of $b, c, s$ quarks and antiquarks, keeping the 1981 parameters and comparing with present experimental data.

\vspace*{5cm}

\vspace*{1.5cm}

\begin{flushleft} CERN-PH-TH/2007-069 \\
April 2007
\end{flushleft}
\vfill\eject
\pagestyle{empty}

\setcounter{page}{1}
\pagestyle{plain}

\section{Introduction}

Everybody knows what was called, in particle physics, the "October revolution" because it took place in
November 1974: the discovery of the $J/\psi$ particle \cite{aaa}, which was rather quickly unerstood
to be a quark-antiquark system, $c\overline c$, where $c$, the charmed quark, was precisely the quark
predicted in 1970 by Glashow, Iliopoulos and Maiani \cite{bb}. It was proposed to treat the system as
a non-relativistic system satisfying the Schr\"odinger equation because the charm quark was heavy, 1
to 2 GeV/$c^2$. Whether this was really allowed or not will not be discussed here. The fact is that,
as we shall see later, it was very successful.

In January 1976, I was invited to attend the "Orbis Scientiae" conference in Coral Gables where I was
planning to give a talk on the diffractive peak at high energy, and on my way, I stopped for a few
days at Rockefeller University. There, the late Baqi Beg told me: "You are an expert on the
Schr\"odinger equation. Can't you explain why all models of Charmonium predict that in between the
ground state $J/\psi$ and what is believed to be the first radial excitation, the $\psi^\prime$, there 
should exist a multiplet of $P$ states". 

I thought about this problem and it is only in 1977 that I found an imperfect answer to this question,
namely a reasonable condition on the quark-antiquark potential which would guarantee this
above-mentioned property. Naturally the existing models satisfied this condition. It should also be
said that the $P$ states were discovered at the right place. It is only much later, in 1984, that with
Baumgartner and Grosse, I found a perfectly natural and beautiful condition to ensure the correct
order of energy levels. But let us come back to the Spring of 1977. I had published one paper on the
level ordering \cite{cc} and a second one on the relative magnitudes of the wave functions at the
origin of the
$J/\psi$ and $\psi^\prime$ \cite{dd}. Professor A. Zichichi had his office exactly opposite to mine.
He read my papers, and thought they were interesting. Then he asked me to give several lectures at the
School of Subnuclear Physics in Erice on \underline{all} aspects of Charmonium during the summer of
1977. I was completely unprepared to do that, and my wife is a witness that I was completely panicked.
The summer came, and the lectures went relatively well \cite{ee}. In the end it turned out that being
forced to give these lectures was a blessing in disguise, a blessing of God, God being personified by
Antonino Zichichi! Indeed in the following years, an important fraction of  my activity was devoted to
Charmonium, or rather Quarkonium, since $b\overline b$ systems had also been discovered, and since
$s\overline s, c\overline s, b\overline s$, could  also be meaningly included among heavy
quark-antiquark systems. Not only mesons but also baryons containing heavy quarks were  studied. 

It is very difficult, both because of the abundance of material and because of the complexity of
certain proofs to present all the results here. I have already said that the theorems proved in 1977
were superseded by the much nicer results of 1984, and so, we will forget about these early theorems.

I would like to explain that later my activity or rather 
"our" activity, because I had many collaborators, was divided into pursuing the \underline { derivation of rigorous
results}  on the energy levels and wave functions of systems satisfying the Schr\"odinger equation (with
possible applications outside quarkonium physics, like atomic physics, muonic  atoms), and of
\underline{phenomenological fits} of the quarkonium spectra which have happened to possess \underline{a very impressive
predictive power}.

Among my collaborators I would list, in chronological order of appearance, Harald Grosse, Reinhold
Bertlmann,  Jean-Marc Richard, Alan Common, Bernard Baumgartner, Joachim Stubbe, Jon Rosner.

Here I want to summarize both aspects of our activity. Concerning rigourous results, there has been
already a Physics Reports published by Grosse and myself \cite{ff} following a Physics Reports by
Quigg and Rosner \cite{ggg} on the same subject. There was also a review presented at the School
of Subnuclear Physics in Erice in 1992 \cite{hh}. Later, Grosse and myself produced a fairly complete
book \cite{jj} recently reprinted in paperback.
This book contains results which concern not only quarkonium but also muonic atoms and alcaline atoms
to which our theorems apply.

In the last section we present an update of the fits of heavy quark systems (quarkonium and heavy
baryons) with a simple model. This is an improved version, containing new experimental data, of a
review at the Montpellier conference in 1997 \cite{kk}.

\section{Order of Energy Levels}

Figures 1 and 2 present the energy levels of the $c\overline c$ and $b\overline b$ systems
respectively. One sees
\begin{itemize}
\item[i)] that the $\ell = 1$ states are between the $\ell =0$ states,
\item[ii)] that the average energy of the $\ell = 1$ states is larger than the average of the energies
of the $\ell = 0$ states immediately above and below, and that the $\ell = 2$ state of the $c\overline
c$ system is \underline{above} the first radial excitation.
\end{itemize}

\begin{figure}[htbp]
\renewcommand{\captionlabeldelim}{.}
\begin{center}
\includegraphics{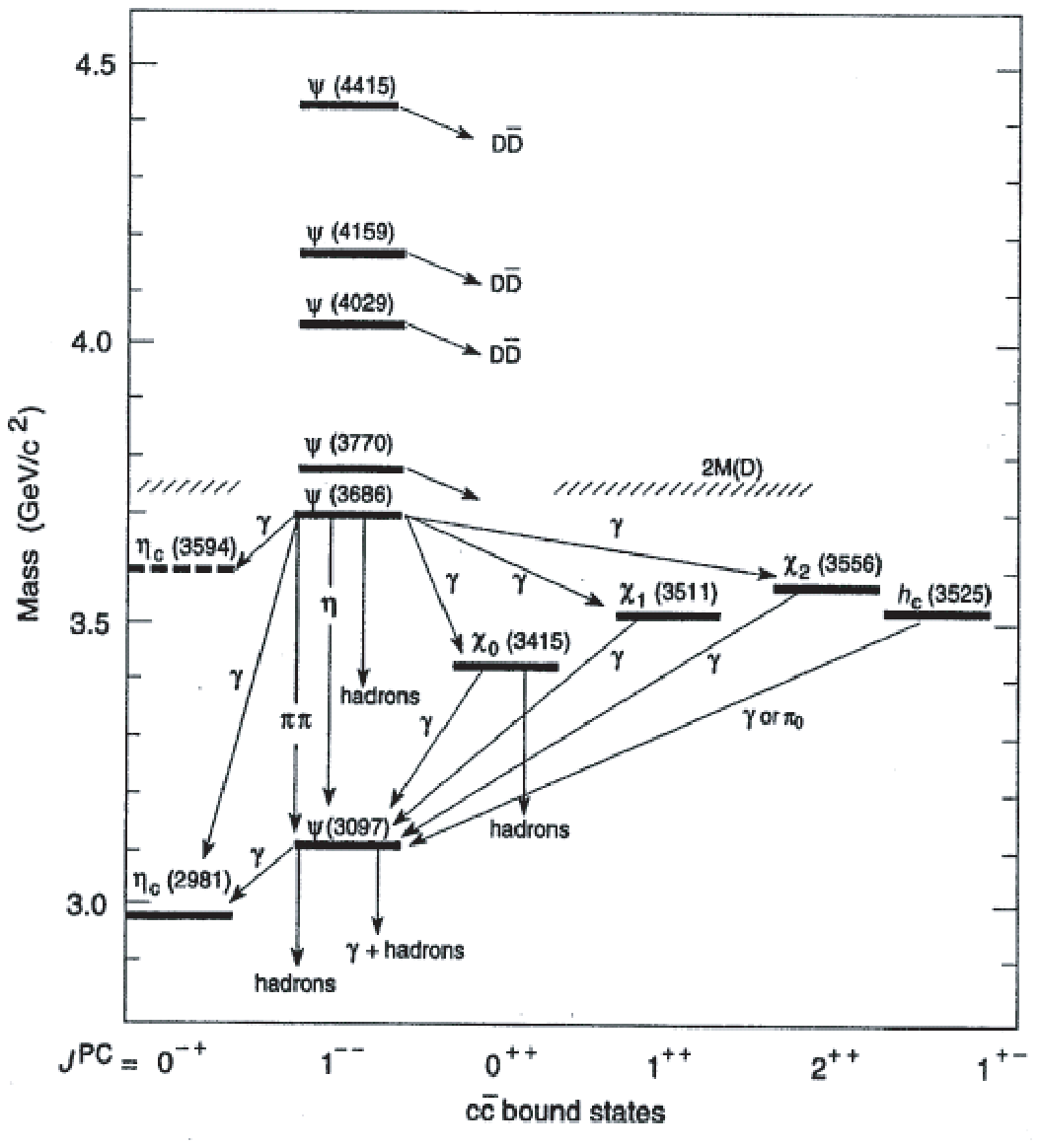}
\end{center}
\caption{}
\end{figure}

\begin{figure}[htbp]
\renewcommand{\captionlabeldelim}{.}
\begin{center}
\includegraphics{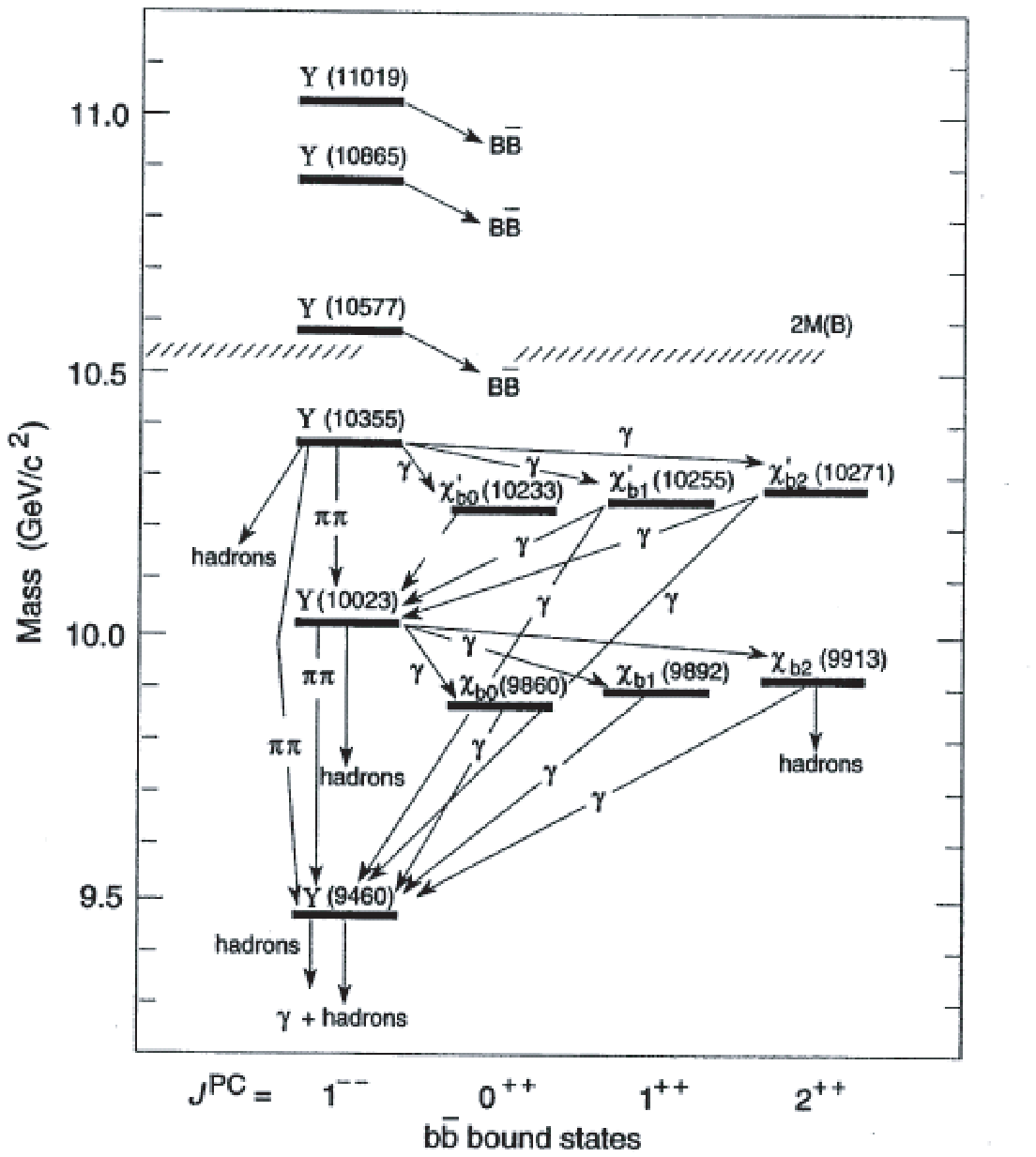}
\end{center}
\caption{}
\end{figure}

 What was found to be the "good" condition to explain i) is that the central potential should be such
that its Laplacian is positive \cite{lll}. 

We can state the following theorem:

\noindent
\underline{Theorem ~I}\\
Let $E(n,\ell )$ be an energy level in a central potential, $n$ being the number of nodes of the
radial wave function, $\ell$ the orbital angular moment. Then  
\beq
E(n+1,\ell) \>< ~ E(n,\ell +1) ~~
\rm{if}~~ r^2 \Delta~~V(r) \>< ~ 0~.
\eeq
Now: why is this a good criteria, for the quark-antiquark potential. The answer is that this is a
strong version of asymptotic freedom.

Call the force between a quark and an antiquark
\beq
-{dV\over dr} = F(r) = - {\alpha(r)\over r^2}
\eeq
In the Coulomb case, $\alpha (r)$ is essentially constant and
\beq
\Delta V = {1\over r^2}~~{d\over dr}~~r^2~{dV\over dr} = 4\pi~~\delta^3(\overline r)
\eeq
i.e., $\Delta V = 0$ outside the origin. 

In QCD, asymptotic freedom can be expressed as
\beq
{d\over dr}~~\alpha(r) > 0~,
\eeq
where $\alpha(r)$ is the running coupling constant and hence $ \Delta V(r) > 0$.

Now let us come to ii). We know, since Newton (according to Markus Fierz), that there are two and only
two kinds of forces for which all bound states have classical closed orbits: the Coulomb and harmonic
oscillator forces.

Theorem I represents a comparison of the potential with the Coulomb potential. It might be interesting
to compare the quark-antiquark potential with the harmonic oscillator potential, $V = r^2$. $V =
a+br^2$ satisfies
$$
{d\over dr}~~{1\over r}~~{dV\over dr} = 0
$$
One can prove the following theorem \cite{mm}.

\noindent \underline{Theorem II}\\
If
\bea
{d\over dr}~~{1\over r}~~{dV\over dr} &\>< ~0~~~~~~~~~~~~~~ \cr
E(n+1,\ell) &\>< ~ E(n,\ell+2)
\eea

In the case of quarkonium, it has been shown by E. Seiler \cite{nn}, from lattice QCD, that the
quark-antiquark potential is concave and increasing. Hence
$$
{d\over dr}~~{1\over r}~~{dV\over dr} = -{1\over r^2}~~{dV\over dr} + {1\over r}~~{d^2V\over dr^2} < 0
$$

In this way, we understand that the mass of the $\psi^"$ $(n = 0, \ell = 2, )$ is higher than the mass
of the $\psi^\prime~(n = 1, \ell = 0)$. We shall come back later on the question of why the $\ell = 1$
states are above the average of the neighbouring two $\ell = 0$ states.

Outside these two theorems on the level ordering there are others, where the criterium is the sign of
$D_\alpha V(r)$ where 
\beq
D_\alpha = {d^2\over dr^2} + (5-3\alpha)~{1\over r}~~{d\over dr} + 2 (1-\alpha)~(2-\alpha)~{1\over r^2}
\eeq
the previous cases correspond to
$$
\alpha = 1 ~~\rm{and}~~\alpha = 2
$$
These theorems are not optimal\footnote{For details, see \cite{jj}, p. 43}, but we have conjectured
with a relatively strong basis that if
\beq
\left( r~{d^2\over dr^2} - (\alpha^2 - 3)~{d\over dr}\right)~V(r) \leq 0
\eeq
with $\alpha > 2$, then $E(n,\ell) <  E(n -1, \ell+\alpha)$, for instance if we take
$$
\begin{array}{ll}
V(r) &= r^4 \\
E(n,\ell) &< E(n-1,\ell+\sqrt{6})
\end{array}
$$
Indeed, for instance
$$
\begin{array}{lll}
E(1,0) &= 11.6,~~E(0,\sqrt{6}) &= 12.6 \\
E(1,5) &= 35,7,~~E(0,5+\sqrt{6})& = 36.3
\end{array}
$$
In another special case, $V(r) = r^5$ a rigorous but painful proof gives 
$$
E(n=0,\ell+\sqrt{7}) > E(n=1,\ell)~.
$$

Let us return now to Theorem I. So far it was applied to quarkonium, but there are other applications.
The first one is muonic atoms. In muonic atoms, the Bohr radius is so small that the muon "sees" the
extension of the nucleus.

Since the nucleus, in the first approximation where it is supposed to be made of protons and neutrons,
has a positive charge distribution, the potential exerted on the muon has a \underline{positive}
Laplacian and
$$
E(n+1,\ell) > E(n,\ell+1)
$$

This is indeed what was found in early calculations of muonic energy levels \cite{oo}. However, this
was in fact independent of the details of the chosen charge distribution. This is seen experimentally.
For instance, using the standard atomic spectroscopic notation, $N,\ell$, total angular momentum,
with $N = n+\ell + 1$, for $(\mu^-~~^{138}Ba)$ atoms \cite{pp}
$$
\begin{array}{ll}
2s_{1/2} - 2p_{1/2} &= +405.41~~ \rm{KeV} \\
3p_{3/2} - 3d_{3/2} &= ~~~~~~ 8.69~~ \rm{KeV} 
\end{array}
$$

One might object that relativistic corrections cannot be neglected. However, it has been shown
\cite{qq}
 (at least for perturbations around the Coulomb potential) that the inequalities survive for the
Dirac equation as long as one compares levels with the same total angular momentum.

The other field where Theorem I is useful is atomic physics, in the approximation where the alcaline
atoms can be treated as one outer electron plus a closed shell. The nucleus, seen from far away by the
outer electron is now essentially pointlike and the effective potential is a Coulomb potential plus
the result of a negative charge distribution. So it has a negative Laplacian. Take for instance
Lithium. It has three electrons. Two of them form a closed shell. What will the outer electron be:
$$
2s~~\rm{or}~2p~~?
$$
It will be $2s$ because of the negativity of the Laplacian. In this way, Theorem I helps to understand
many aspects of the Mendeleiff classification which, previously, were explained by not very
convincing handwaving arguments. Figure 3, taken from \cite{rr} shows that these considerations hold
also for excited states of the lithium atom. 

\begin{figure}[htbp]
\renewcommand{\captionlabeldelim}{.}
\begin{center}
\includegraphics{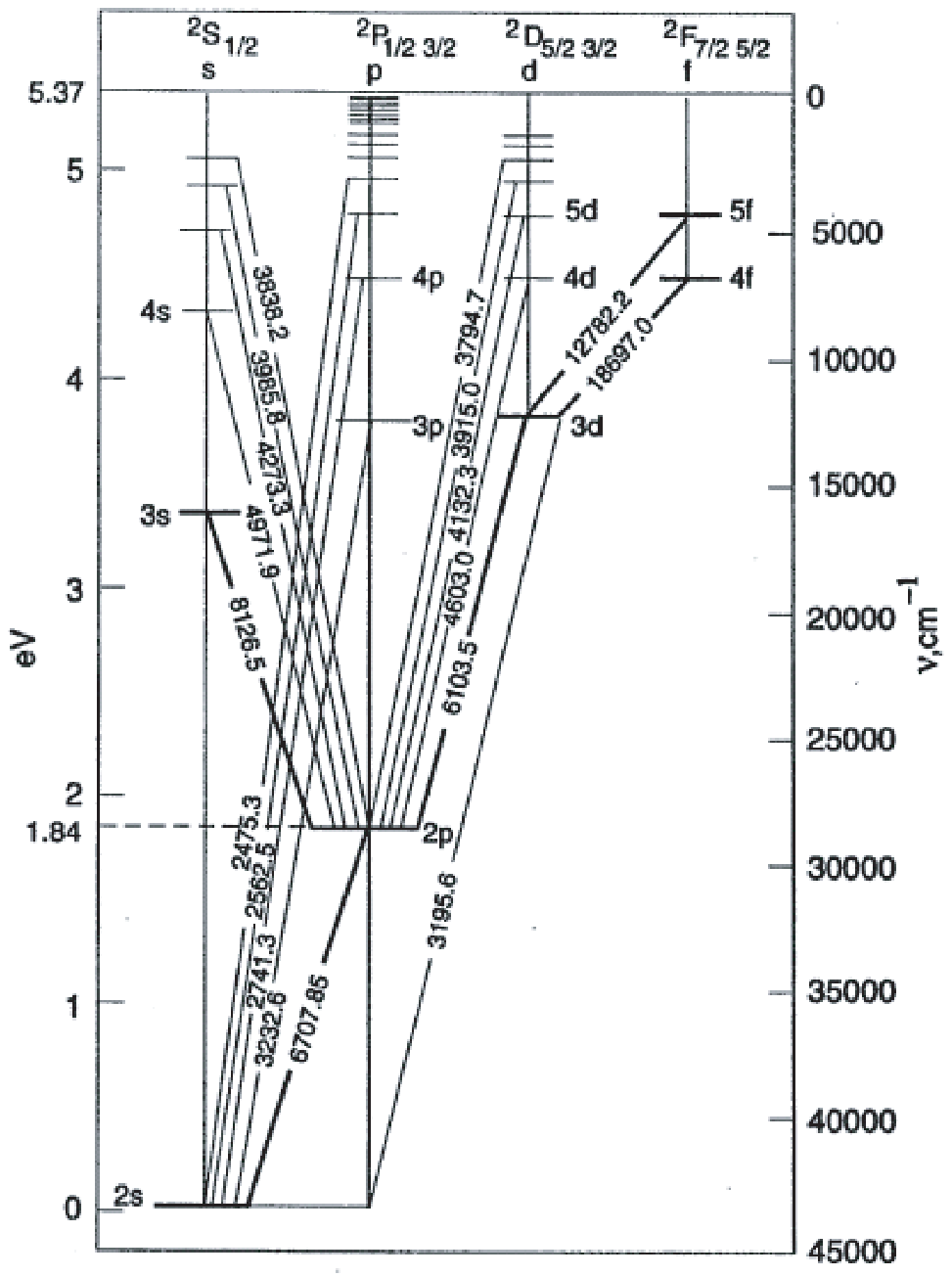}
\end{center}
\caption{}
\end{figure}

\newpage

One sees also that for large principal quantum numbers
the Coulomb degeneracy tends to reappear. On Fig. 3, it is interesting to see that
$$
3s < 3p < 3d
$$
more precisely, one has, theoretically \cite{ss}, and experimentally \cite{tt}:
\bea
&&3s_{1/2}~~~~~~ < 3p_{1/2}~~~~~~~ < 3p_{3/2}~~~~~~~ < 3d_{3/2}~~~~~~ < 3d_{5/2} \cr
&&27206.066 < 30925.517 < 30925.613 < 31283.018 < 31283.053~~\rm{cm}^{-1} 
\eea

These inequalities are valid for an outer electron satisfying the Dirac equation with a monotonous
increasing potential such that $\Delta V < 0$. The assumption of monotonicity is probably superfluous.

\section{Spacing of Energy Levels}

Since energy levels depend on two quantum numbers, $n$ and $\ell$, the problem is very broad and very
difficult to explore completely. We shall restrict ourselves to two extreme cases:
\begin{itemize}
\item[a)] Compare the spacing of the energy levels for fixed $n$, and in fact we shall restrict
ourselves to $n = 0$, i.e., the purely angular excitations.
\item[b)] Study the relative spacings of the energy levels for fixed $\ell$, specifically $\ell = 0$.
\end{itemize}

While, for case a) there is a large amount of exact results, for case b) we have only imperfect
indications. The results obtained in a), however, will allow us to answer question ii) in Section~II.

In these problems a simple reference potential is the harmonic oscillator $V = r^2$. There all the
fixed $n$, increasing $\ell$ spacings are equal and all the fixed $\ell$ increasing $n$ spacings
are equal.

\noindent a) Spacing of purely angular excitations $n = 0$, increasing $\ell$. There one gets easily

\noindent \underline{Theorem~III} \cite{uu}\\
If
\bea
{d\over dr}~~{1\over r}~~{dV\over dr} < 0 \cr \cr
{E(0,\ell+2) - E(0,\ell+1)\over E(0,\ell+1) -E(0,\ell)} < 1
\eea
There exists also a differential form of this inequality. 

If we combine this with Theorem II, which gives, under the same conditions
$$
E(1,\ell) < E(0,\ell+2)
$$
we get
$$
E(0,\ell+1) > {1\over 2}~\left[E(0,\ell) + E(1,\ell)\right]
$$
and the special case
$$
E(0,1) > {1\over 2} \left[E(0,0) + E(1,0)\right]
$$
which is precisely what is observed in the $c\overline c$ and $b\overline b$ spectra.

We would also like to have the analogue inequality for the second $\ell = 1$ state, $E(1,1)$ but so
far this is not the case.

A more subtle inequality for angular excitations is obtained if $\Delta V(r) \>< ~ 0$. Denote
$E(0,\ell)$ as $E_\ell$, then

\noindent
\underline{Theorem IV} \cite{vv}\\
If 
\bea
{d\over dr}~r^2~{dV\over dr}~\>< ~ 0~~, \cr \cr
{E_{\ell+1} - E_{\ell}\over E_\ell - E_{\ell-1}} ~{\><}~ {2\ell+3\over 2\ell + 1}~~\left({\ell\over
\ell + 2}\right)^2~,
\eea
where the right-hand-side corresponds to the pure Coulomb case. Application to the $c\overline c$
system is rather frustrating. We get
$$
{E_2 - E_1\over E_1 - E_0} > {5\over 27}~,
$$
trivially satisfied by experiment.

Muonic atoms, on the other hand, are much more interesting. They also have $r^2 \Delta V > 0$ and the
spacing between purely angular excitations is what is most easily observed by looking at cascading
$\gamma$ rays by favoured electric dipole transitions. For small nuclear charge $Z$ or for large
angular momenta the deviations for the Coulomb value are small while they are large for small $\ell$
and large $Z$.

After relativistic corrections, we get \footnote{The references on the experimental material used in
Tables 1, 2, 3 and 4 can be found in Ref. \cite{vv}.}

{\small
\begin{center}
Table 1 \\
\vglue0.3cm
\begin{tabular}{|l l l l  | }
\hline
  $Z$    & ${E_{3d}-E_{2p}\over E_{2p} - E_{1s}}$& 
${E_{4f}-E_{3d}\over E_{3d} - E_{2p}}$ & 
${E_{5g}-E_{4f}\over E_{4f} - E_{3d }}$ \\
\hline
$Z\rightarrow 0$ & 0.185     & 0.350 & 0.463 \\
20  & 0.202   & 0.350   & 0.463  \\
40  & 0.253   & 0.350   & 0.463  \\
60  & 0.333	  & 0.353   & 0.463  \\
80  & 0.462   & 0.370   & 0.463   \\  
\hline
\end{tabular}
\end{center}}

Conversely, for Alcaline atoms we have $\Delta V < 0$. Here we just give the
\underline{Lithium~sequence}, i.e., ions with three electrons and increasing charge. Here, it is just
the reverse. When $Z$ gets large, the Coulomb limit is approched because the charge of the nucleus
dominates on the electron cloud. For extremely large $Z$ (this is fictitious) the orbits are inside
the cloud and do not see it.

{\small
\begin{center}
Table 2 \\

\vglue0.3cm
\begin{tabular}{|l l l  | }
\hline
  $Z$    & ${E_{4f}-E_{3d}\over E_{3d} - E_{2p}}$& 
${E_{5g}-E_{4f}\over E_{4f} - E_{3d}}$ \\
\hline
~3  & 0.326   & 0.462 \\
~6  & 0.329   & 0.462  \\
~9  & 0.334   & 0.462  \\
12  & 0.338	  & 0.462  \\
15  & 0.339   &    \\
$\infty$  & 0.350   & 0.463    \\ 
\hline
\end{tabular}
\end{center}}

Finally we give inequalities on the spacing between states with the same quantum numbers and different
total angular momentum, i.e., the fine splitting. If, keeping $n = 0$
\beq
\delta = E(\ell , J \equiv \ell + {1\over 2}) - E(\ell , J^\prime = \ell - {1\over 2})
\eeq
we have in the semi-relativistic approximation
\beq
\delta = {2\ell + 1\over 4m^2} \left< {dV\over dr}\right>
\eeq
where the expectation value is to be taken using the \underline{Schr\"odinger} wave function.

We have two theorems \cite{vv}.

\noindent \underline{Theorem V}
\beq
\delta(\ell) \<> ~ \Delta(\ell) = {2\over
m \ell}~~{(\ell+2)^4\over (2\ell+3)^2}~~\left(E_{\ell+1}-E_\ell\right)^2
\eeq
and
\beq
{\delta (\ell +1)\over \delta (\ell )} \>< ~ {\ell (\ell + 1)^3\over (\ell + 2)^4}~~,~~~\rm{if}~~~
r^2 \Delta V(r) \>< ~ 0
\eeq
for \underline{muonic~atoms}, i.e., $r^2 \Delta V(r) > 0$, we give only a sample of the results and
forget the uncertainties to improve legibility

{\small
\begin{center}
Table 3 \\
\vglue0.3cm
\begin{tabular}{|l l l l l | }
\hline
  $Z$    & $\delta (1)$ & $\Delta (1)$ & $\delta (2)$ & ${\delta (2) \over \delta (1)} > 0.098$ \\
									& in keV &&&\\
\hline
26 & ~~4.2   & ~~4.4  & 0.47 & 0.112 \\
33 & 11.10  & 11.45   & 2.00 & 0.18 \\
41 & 23.15  & 29.10   & 2.68 & 0.116\\
50 & 45.7	  & 60.60   & 5.65 & 0.123 \\
  \hline
\end{tabular}
\end{center}}

Applications to Alcaline atoms, with $\Delta V < 0$, are on the other hand rather
disappointing. Let us remember that Eq. (12) is violated by the levels of the Sodium atom since
the sign of the spacing of the well-known Sodium doublet (which produces by transition to the ground
state the horrible yellow light) is the opposite. Exact treatment by the Dirac equation does not help.
It is a typically many-body effect.

Concerning the lithium isoelectronic sequence, i.e., ions with three electrons, the situation is not
as bad in the sense that $\delta (1)$ and $\delta (2)$ have the right sign. However, the inequality
\beq
\delta(1) > \Delta (1)~~\rm{and}~~ \delta (2 > \Delta (2)
\eeq
are violated for Lithium itself. It is only for $Z \geq 6$ (Be II, etc.) that they are satisfied. This
is seen in Table 4 (the units are cm$^{-1}$)

{\small
\begin{center}
Table 4 \\
\vglue0.3cm
\begin{tabular}{|l l l l l l | }
\hline
  $Z$    & $\delta (1)$ & $\Delta (1)$ & $\delta (2)$ & $\Delta (2)$& ${\delta (2) \over \delta (1)} <
0.098$\\
									& in cm$^{-1}$ &&&&\\
\hline
~3 & ~~0.337   & ~~0.422  & ~~~~0.037 & ~~~0.036  &  0.1097 \\
~5 & ~34.1     & ~34.2    & ~~~~3.1   & ~~~2.5 			&  0.0909\\
~6 & 107.1     & 106.6    & ~~~10.5   & ~~~9.278		&  0.098\\
~9 & 975.8	    & 962.7    & ~~~90.0   & ~~87 					&  0.092\\
12 & 3978      & 3938     & ~~470     & 362       & 0.0118 \\
15 & 11310     & 11100    & 1000      &           & 0.088 \\
24 & 90910     & 90000    &           &           &       \\
  \hline
\end{tabular}
\end{center}}

We see also that $\delta(1) / \Delta (1)$ is always very close to unity. There are also inequalities
of the same kind for hyperfine splittings but, for these, we refer to the original work \cite{vv}.

We turn now to the problem of spacings between $\ell = 0$ levels \cite{ww}. We present a number of
incomplete results, which give strong indications, but completely rigorous proofs are lacking. Two
things are certainly true:
\begin{itemize}
\item[a)] the $\ell = 0$ levels of the harmonic oscillator potential are equally spaced, as everybody
knows;
\item[b)] the $\ell = 0$ levels of the linear potential have a spacing decreasing with $n$.
\end{itemize}

We study now the neighbourhood of a), i.e., potentials close to the harmonic oscillator potential.

A preliminary remark is that the spacing between the $n = 0$ levels will remain constant not only if
we replace $r^2$ by $A~r^2 + B$, but also if we add a $C/r^2$ term. Adding such a term is equivalent
to changing the angular momentum by an (generally not integer) amount $\ell$ such that $C = \ell (\ell
+1)$. Since the ``Regge trajectories" of the harmonic oscillator are linear and parallel, the spacing
between the energy levels will not change.

An interesting quantity to control the spacings is
\beq
Z (V,r) = {d\over dr}~~r^5~~{d\over dr}~~{1\over r} ~~{dV\over dr}
\eeq
$Z$ vanishes for 
\beq
V = \rm{const.},~~ V = r^2~,~~V = {1\over r^2} ~.
\eeq

What we have proved, using a new kind of raising and lowering operators is that if
\beq
 V = r^2 + \lambda v~~,
\eeq
 for $\lambda$ sufficiently \underline{small}, the spacing between the energy levels
\underline{increases} with $n$ if $Z(V,r)>0$ $\forall r$ and the spacing between the energy levels
\underline{decreases} with $n$ if $Z(V,r) < 0$ $\forall r$.

However, in the non-perturbative case, we do not know what happens. In fact we even have
counter examples.

Take the ``partially soluble" potential \cite{yy}
$$
V = r^6 - 9 r^2
$$
the first five energy levels are given by
\beq
\left\{
\begin{array}{ll}
E_0, E_4 &= \mp \sqrt{480 + 96 \sqrt{11}} \\
E_1, E_3 &= \mp \sqrt{480 - 96 \sqrt{11}} \\
E_2 &= 0
\end{array} 
\right. 
\eeq

Obviously, even though $Z(V,r)$ is positive, the levels show a symmetric pattern around $E_2 = 0$.
However, there is no known counterexample in which $V$ is \underline{monotonous} increasing.

We are tempted to make the conjecture that if $Z > 0$ and if the potential is monotonous, the spacings
increase, and if $Z < 0$ and the potential is monotonous, the spacings decrease. ``Experimental" tests
with $V = r^\alpha$ indicate that the spacing increases with $n$ if $\alpha > 2$ and decreases with
$n$ if $\alpha < 2$.

We study now the neighbourhood of b), the purely linear potential. Then the solutions of the
Schr\"odinger equation are Airy functions. Then, the energy levels are given by $E_n = -r_{n-1}$,
where $r_n$ is the $n^{th}$ zero of the Airy function. It is obvious that the spacing between $r_n$
and $r_{n+1}$ is smaller than the spacing between $r_{n-1}$ and $r_n$ because the potential is
stronger.

We strongly believe that for all concave potentials, which correspond to the physical situation for
quark-antiquark systems \cite{nn}, the spacing between $\ell = 0$ levels decreases with $n$. This is
precisely what is indicated by the WKB approximation, in which we can regard $n$ as a continuous
variable. The spacing will decrease if 
$$
{d^2E\over d~n^2} < 0~,~~~{\rm{i.e.}}~~~ {d^2n\over dE^2} > 0~.
$$ Indeed, in the WKB approximation
\bea
{d^2n\over dE^2} = &{1\over 2\pi}~~{1\over V^\prime (0) \sqrt{E-V(0)}}\cr
& - {1\over 2\pi}~~\int_{E-V>0}~~{V^"\over V^{\prime 2}\sqrt{E-V}}~~dr
\eea
which is positive if $V^\prime (0) > 0$ and $V^" < 0$.

The $b\overline b$ spectrum possesses this property, at least below the $B\overline B$ threshold
\cite{zz}:
$$
\begin{array}{ll}
M_{\gamma ^\prime} - M_\gamma &= 560 ~\rm{MeV}\\
M_{\gamma ^"} - M_{\gamma ^\prime} &= 332~\rm{MeV}
\end{array}
$$

\section{The Wave Function at the Origin, the Kinetic Energy, etc.}

Perhaps it is worth mentioning results on the wave function at the origin because one of the two
papers which initially attracted the attention of Professor Zichichi is precisely on this subject
\cite{dd}.

The wave function at the origin appears in the so-called Van Royen-Weisskopf formula \cite{aai} (also
proposed by Pietschmann and Thirring, M. Krammer and H. Krasemann) which controls the leptonic width of
quarkonium:
\beq
\Gamma_{e^+e^-} = 16 \pi \propto e^2_Q~~{|\psi (0)|^2\over M^2}~,
\eeq
where $e_Q$ designate the charge of the quarks in units of $e$. $|\psi (0)|^2$ also controls the
hadronic width, i.e., the decay into three gluons of quarkonium. The reduced wave function at the
origin is given for the $\ell = 0$ wave function by a formula attributed to Schwinger:
\beq
\left(u^\prime_n(0)\right)^2 = 2m \int^\infty_0~~{dV\over dr}~~u^2_n~~dr~,
\eeq
where
$$
u = {r\psi\over \sqrt{4\pi}}~,
$$
with
$$
\int~~u^2_r~~dr = 1
$$
if $V$ is \underline{linear}, ${dV\over dr} =$ const, and hence, because of the normalization of
$u_n$, $|u^\prime _n(0)|^2$ and $|\psi_n(0)|^2$ are independent of $n$.

We proved the following theorem:  if $V$ is concave, i.e., if $V^" < 0$
\beq
|\psi_1(0)|^2 < |\psi_0(0)|^2~~,
\eeq
 if $V$ is convex, we get the reverse.

Indeed, even if you take into account the change of mass in the Weisskopf-Van Royen formula,
experiment indicates that the leptonic widths of the $J/\psi$  and $\psi^\prime$ fit with a concave
potential, which is what we expect from lattice QCD \cite{nn}.

Can we go beyond that result? All we can say is that we have strong indications that the leptonic
widths decrease with $n$ if $V$ is concave. For instance we can prove that $|\psi_n(0)|^2$ goes to
zero for $n \rightarrow \infty$ if $V$ is concave.

Outside the wave function at the origin there are other quantities of interest on which we have a
control, such as the mean kinetic energy, the root mean square radius, the electric dipole transition
matrix elements, etc. For all these we send the reader to Ref. \cite{jj}. Let us just mention that this
allows to set constraints on the ``Schr\"odinger mass" of the quarks, if the potential is flavour
independent, from the values of the $\ell = 0$ and $\ell = 1$ energy levels of the $c\overline c$ and
$b\overline b$ systems:
\beq
m_b - m_c > 3.29~\rm {GeV}~,
\eeq
while a na\"\i ve approach gives
\beq
m_b - m_c > {1\over 2}~\left(M_{b\overline b} - M_{c\overline c}\right) = 3.18 ~\rm {GeV}
\eeq
The mean kinetic energy of the system, $T$, that we call $T(m)$, where $m$ is the
common quark mass in quarkonium, appears in the change in binding energy of a quark-antiquark
system when the mass changes for a fixed potential (called flavour-independent for quarkonium), which
is, from the Feynman-Hellman theorem
\beq
{d\over dm}~~E(m) = -{T(m)\over m}
\eeq
This proves already (25), from the positivity of $T$.

Now, with $M > m$
$$
E(M) = E(m) - \int_m^M~~{T(\mu )\over \mu}~~d\mu
$$
From the concavity with respect to any parameter entering linearly in the Hamiltonian, we get that $mT(m)$ is
increasing with $m$. In fact, for a square well potential $mT(m)$ is constant.

However, if the potential is concave, which we believe from lattice QCD \cite{nn}, we get a stronger
result \cite{bbi}:
\beq
m^{1/3} T(m) {\bf{\nearrow}}
\eeq
with $m$.

On the other hand, an inequality on $T(m)$ has been obtained \cite{cci}
\beq
T(m) > {3\over 4}~~\left[ E(n=0, \ell = 1, m) -E (n=0, \ell = 0, m \right]~,
\eeq
which is saturatedby a harmonic oscillator potential.

From these considerations it is possible to get a lower limit to the mass difference between the $b$
quark mass and the $c$ quark mass, taking into account that the mass of a $Q\overline Q$ system is
$2m_Q + E(m_Q)$:
\beq
m_b - m_c > 3.32~\rm{MeV}
\eeq
Without (27), we would get the weaker result (24).
This is only a sample of the many results obtained in this domain, for which we send the reader to
Ref. \cite{jj}.

\section{A Fit of Heavy Quark Systems by a Potential Model}

Now, we are leaving mathematical physics and turning to (dirty) phenomenology. After the discovery of
the Upsilon and Upsilon prime systems by Lederman in 1977 \cite{ddi}, Quigg and Rosner made the remark
that the spacing of the levels of the $c\overline c$ and $b\overline b$ systems is almost the same. If
a potential model is acceptable, and if this potential is flavour independent, it is tempting to say
that the potential could just be logarithmic, i.e., $C \ell n~ r$. Then it is almost obvious that the
spacings of all energy levels will be independent of the mass, because re-scaling $r$ to take into
account the change of the mass will shift the potential by a constant. In 1981, I tried a small
generalization of this by taking $V = A + Br^\alpha$, and adjusting $A, B$ and $\alpha$ to the known
levels. I added too a Fermi-like hyperfine splitting to ``explain" the $J/\psi - \eta_c$ mass
difference. This seemed to be successful \cite{eei}, \cite{ffi}, for mesons made of $b, c$ and even $s$
quarks and antiquarks. At the same time, Jean-Marc Richard, using the ``rule" $V_{QQ} = {1\over
2}~V_{Q\overline Q}$ reproduced beautifully the mass of the $\Omega^-$ baryon \cite{ggi}.

Then, my model was taken with disbelief by many physicists, including some friends, especially
Russians, who could not understand how a na\"\i ve potential model could be justified. Most people
preferred QCD sum rules, or lattice QCD, still in infancy.

However, I took the position that it was worth experimenting with this model even if it was difficult
to justify. With time, many new particles and energy levels were discovered and happened to fit
perfectly with this model. I remember a seminar by Lorenzo Foa, announcing the discovery of the $B_s$
meson by Aleph at LEP and saying that it was not necessary to give its mass because I had already
predicted it. With, may be, one exception, this continues till now.

In fact, a first version including only $b$ and $c$ quarks was proposed \cite{eei}. However, seeing the
incredible success of the fit, Murray Gell-Mann suggested that one should go further in the ``absurd"
(since $c$ quarks are not really non-relativistic) and include the strange quark. Let me give the
numerical elements of the fit, made in 1981 \cite{ffi}, with the existing experimental information. The
potential is
$$
V = A + Br^\alpha
$$
the fit gives
$$\begin{array}{ll}
A & = -8.064\\
B & = ~~~6.870~,
\end{array}
$$
in units which are powers of GeV
$$
\alpha = 0.1
$$
$$
\begin{array}{
ll}
m_b &= 5.174\\
m_c &= 1.8 \\
m_s &= 0.518
\end{array}
$$
(notice that this agrees with the lower limit (29) on $m_b-m_c$).

Futhermore, the large hyperfine splitting between $J/\psi$ and $\eta_c$ has to be taken into account
by a phenomenological spin-spin interaction
$$
{C\over m_1m_2}~~(\sigma_1.\sigma_2)~~\delta^3(\vec r_1 - \vec r_2)
$$
$C$ is adjusted to the $J/\psi - \eta_c$ splitting.

Overall this is a \underline{seven} parameter fit. We do not try to predict the fine splitting between the $\ell =
1$ states and just predict the weighted average.

Table 5 gives the results. It contains 30 experimental numbers.  It is an updated version of a table presented at the Montpellier 1997
conference \cite{kk}. The figures with stars are experimental results which were not known in 1981 (ten
stars!).  The fact that the $\eta^\prime_c$ is 15 MeV higher than predicted has been understood long ago because of the vicinity of the $D \bar D$ threshold (by J.-M. Richard).

\newpage

\begin{center}
Table 5 \\
Masses and relative leptonic widths for $c\overline c, b\overline b, s\overline
s$,
\\
$c\overline s, b\overline s, b\overline c$, and of the $\Omega$ and $\Omega_c$
baryons.\\
\vglue0.3cm
\begin{tabular}{|l l l l l l | }
\hline
  Quark    & States & Theory &  Experiment & Theory & Experiment \\
System &&&&& \\
\hline
$c\overline c$ & $J/\psi$      & 3.095 & 3.097 & 1    & 1 \\
& $\psi^\prime$ & 3.687 & 3.686 & 0.40 & 0.46 $\pm$ 0.06\\
& $\psi^{\prime \prime\prime}$ & 4.032 & 4.040 & 0.25 & 0.16 $\pm$
0.02
\\
               & $\chi_c$ average & 3.502 & 3.525 && \\
               & and $h_c$ &&&& \\
               & $\psi^"$ average & 3.787 & 3.770 && \\           
              & $\eta_c $& 2.980 & 2.980 & & \\
              & $\eta^\prime_c$ & 3.641 & 3.656$^*$&&
\\
&&&&&\\
$b\overline b$ & $\Upsilon$   & ~9.46 & ~9.46 & 1 & 1 \\
               & $\Upsilon^\prime$ & 10.02& 10.02 &  0.35 & 0.32 $\pm$ 0.05 \\
               & $\Upsilon^"$ & 10.36 & 10.35 & 0.27 & 0.24 $ \pm$ 0.05 \\
&$\Upsilon^{\prime\prime\prime}$ & 10.60 & 10.58 & 0.21 & 0.23 $\pm$ 0.06 \\
&$\Upsilon^{iv}$              & 10.76 & 10.86 $^*$ &  &  \\
&$\chi^b$ average & ~9.86 & ~9.90 $^*$ && \\
& $\chi^\prime _v$ & 10.24 & 10.26 $^*$ && \\
&&&&&\\
$s\overline s$ & $\phi$ & 1.02 & 1.02 && \\
& $f_1$ average & 1.42 & 1.44 && \\
& P state &&&&\\
& $\phi^\prime$ & 1.634 & 1.680 && \\
&&&&&\\
$c\overline s$ & $D_s$ & ~1.99 & ~1.97 && \\
& $D^*_s$ & ~2.11 & 2.11 $^*$ && \\
& $D^{**}_s$ average & 2.537 & 2.536 $^*$ && \\
&&&&&\\
$b\overline s$ & $B_s$ & 5.354 & 5.369 $\pm$ 0.005 $^*$ && \\
& $B^*_s$ & 5.408 & 5.416 $\pm$ 0.004 $^*$ && \\
&&&&&\\
$b\overline c$ & $B_c$ & ~6.25 & 6.4$\pm$ 0.39 $\pm$ 0.13 $^*$ && \\& $B^*_c$ &
~6.32 &&&\\
&&&&&\\
$\Omega$ && ~1.666 & ~1.672 && \\
&&&&&\\
$\Omega_c$ && ~2.708 & ~2.697 $\pm$ 0.0025$^{*)}$ && \\
\hline
\multicolumn{6}{|c|}{
$^{*)}$ These experimental numbers have been obtained \underline{after} the initial
fit in 1981.}\\
\hline
\end{tabular}
\end{center}

\newpage

We turn now to the baryon sector. Here we have been using the rule $V_{QQ} = {1\over 2}~V_{Q\overline
Q}$. Why this rule? A three quark system must be colour singlet. Hence in a three quark baryon, a
quark pair must be a $\overline 3$ colour state. So, if two quarks are close together, the third quark
sees them as a $\overline 3$ state, i.e., an antiquark. Hence the potential between the third quark
and the quark pair is $V_{Q\overline Q}$. Dividing by two we get a potential ${1\over 2}~V_{Q\overline
Q}$ between the third quark and each of the quarks of the pair.

This may seem doubtful but worth trying. Jean Marc Richard has done this for the $\Omega$ particle
made of three strange quarks \cite{ggi}. This gives 
$
M_\Omega = 1666$ MeV,
 to be compared with the experimental value 1672 MeV. Encouraged by this result
we predict \cite{hhi}
$$
M_{\Omega_c} = 2708 ~\rm{MeV},
$$
which is now, experimentally \cite{jji} 2697.5 $\pm$ 2.5 MeV, only 10 MeV away.

Other predictions, still to be tested, are
$$
\begin{array}{ll}
M_{\Omega_{cc}} = 3.737 ~\rm {MeV}\\
M_{\Omega_{ccc}} = 4.787 ~\rm {MeV}
\end{array}
$$

The $ccc$ system, according to Bjorken \cite{kki}, is one of the most interesting quark systems. It is
stable with respect to strong interactions and has a lifetime of the order of $2\times 10^{-13}$
seconds. It might be difficult to produce, but, who knows, it might be seen at the LHC. After all, at
LEP, many interesting particles such as $B_s, B_c$ or $\Lambda_b$ were seen while their observation
was not initially planned.
\vspace{2cm}

\noindent
{\bf ACKNOWLEDGEMENTS}

I have already mentioned, at the beginning,  the crucial role of A. Zichichi, who, by asking me to
lecture on quarkonium in Erice in 1977 induced me to investigate a whole field of research. This, I
believe, turned out to be very fruitful. Among all those to whom I am grateful, I would like to single
out, on the mathematical physics side, H. Grosse who co-signed with me the book "Particle Physics and
the Schr\"odinger Equation", and also R. Bertlmann, A.K. Common and J. Stubbe, and on the
phenomenological side, J.M. Richard. I am grateful to many other physicists (some are no longer with
us): B. Baumgartner, M.A.B. B\'eg, J.S. Bell, R. Benguria, Ph. Blanchard. K. Chadan, T. Fulton, V.
Glaser, A. Khare, J.D. Jackson, R. Jost, P. Landshoff, H. Lipkin, J.J. Loeffel, J. Pasupathy, C.
Quigg, T. Regge, J. Rosner, A. De R\'ujula, A. Salam, P. Taxil.

Finally, I would like to thank my wife, Schu, for her strong moral support during the preparation of
the 1977 lectures, and for her insistance to convince me to write a book with H.~Grosse; I also
thank M.N. Fontaine, who, though retired, typed this article because my brain is too rusty to learn
Latex.

\end{document}